\documentclass[rmp,amsfonts,showpacs,showkeys]{revtex4}
\usepackage{graphicx}
\RequirePackage{times}
\RequirePackage{mathptm}

\begin{document}

\title{Physics and metaphysics looks at computation}

\author{Karl Svozil}
\email{svozil@tuwien.ac.at}
\homepage{http://tph.tuwien.ac.at/~svozil}
\affiliation{Institut f\"ur Theoretische Physik, University of Technology Vienna,
Wiedner Hauptstra\ss e 8-10/136, A-1040 Vienna, Austria}

\begin{abstract}
As far as algorithmic thinking is bound by symbolic paper-and-pencil operations,
the Church-Turing thesis appears to hold.
But is physics, and even more so, is the human mind, bound by symbolic paper-and-pencil operations?
What about the powers of the continuum, the quantum, and what about human intuition, human thought?
These questions still remain unanswered.
With the strong Artificial Intelligence assumption, human consciousness is
just a function of the organs
(maybe in a very wide sense and not only restricted to neuronal brain activity),
and thus the question is relegated to physics.
In dualistic models of the mind, human thought transcends symbolic paper-and-pencil operations.
\end{abstract}


\maketitle

\tableofcontents

\section{Computation is physical}

It is not unreasonable to require
from a ``useful'' theory of computation that any
capacity and feature of physical systems (interpretable as ``computing
machines'') should be reflected therein and
{\it vice versa.}
In this way, the physical realization confers power to the formal method.

Conversely, the formalism might ``reveal'' some ``laws'' or structure in the
physical processes.
With the Church-Turing thesis, physics also acquires a definite, formalized
concept of ``physical determinism'' as well as ``undecidability,'' which is lacking
in pre-Church-Turing times.
Indeed, the Church-Turing thesis can be perceived as part of
physics proper, and its assumption be interpreted as indication that
the Universe amounts to a huge computational process;
a suspicion aleady pursued by the Pythagoreans.
Such perception does not fix the lapse of evolution entirely; in a Laplacian-type
monotony, but still allows for dualism and ``miracles''
through the influx of information from interfaces, as will be discussed below.

The recognition of the physical aspect of the Church-Turing thesis---the
postulated equivalence between the informal notion of ``algorithm,''
and recursive function theory as its formalized
counterpart---is not new
\cite{scilard-col,brillouin1,brillouin2,maxwell-demon,rogers1,odi:89,pit:90,galindo-02}. In particular
Landauer
has pointed out on many occasions
that computers are physical systems, that computations are
physical processes and therefore are subject to the laws of physics
\cite{landauer:61,landauer-67,landauer:82,landauer-87,landauer-88,landauer-89,landauer,landauer-94,landauer-95}.
 As Deutsch puts it \cite[p. 101]{deutsch},
 \begin{quote}
 {\em
 ``The reason why we find it possible to construct, say, electronic
 calculators, and indeed why we can perform mental arithmetic, cannot
 be found in mathematics or logic. {\em
 The reason is that the laws of physics `happen to' permit the
 existence of physical models for the operations of arithmetic}
 such as addition, subtraction and multiplication.
 If they did not, these familiar operations would be
 noncomputable functions. We might still
 know {\em of} them and invoke them in mathematical proofs
 (which would presumably be called `nonconstructive') but we could
 not perform them.''
 }
 \end{quote}
One may indeed perceive a strong interrelationship between the way we do
mathematics, formal logic, the computer sciences and physics. All these
sciences have been developed and constructed by us in the context of our
(everyday) experiences.
The Computer Sciences are well aware of this
connection. See, for instance,
Odifreddi's review
\cite{odi:89}, the articles by
 Rosen \cite{rosen} and Kreisel \cite{kreisel},
 or
 Davis'
 book \cite[p. 11]{davis-58}, where the following question is
asked:
 \begin{quote}
 {\em `` $\ldots$ how can we ever exclude the possibility of our
 presented,
 some day (perhaps by some extraterrestrial visitors), with a (perhaps
 extremely complex) device or ``oracle'' that ``computes'' a
 noncomputable function?''
 }
 \end{quote}

In what follows, we shall briefly review some aspects of the interrelationship
between physics and computation.
We believe that it is the nature of the subject itself which prevents a definite
answer to many questions, in particular to a ``canonical'' model of computation
which might remain intact as physics and the sciences evolve.
So, we perceive this review as a snapshot about the present status
of our thinking on feasible computation.

\section{Paper-and-pencil operations in physics}

After Alonzo Church \cite{church30,church36} conceptualized an equivalent notion of ``effective computability''
with an ``Entscheidungsproblem'' (decision problem) in mind
which was quite similar to the questions G\"odel pursued in Ref.~\cite{godel1},
Alan Turing \cite{turing-36} enshrined that part of mathematics, which can be ``constructed''
by paper and pencil operations, into a Turing machine which possesses a potentially
unbounded one-dimensional tape divided into cells,
some finite memory and some
read-write head which transfers back and forth information from the tape to this memory.
A table of transition rules figuring as the ``program'' steers the machine deterministically.
The behaviour of a Turing machine may also be determined by its initial state.

Furthermore, a universal Turing machine is capable
of simulating all other Turing machines (including itself).
According to Turing's definition stated in Ref.~\cite{turing-36},
a number is computable if its decimal can be written down by a machine.
In view of the ``algorithm'' created by Chaitin \cite{chaitin3,calude:02} to ``compute''
the halting probability and encodable by almost every conceivable programming language such as C or Algol,
one should add the proviso that any such Turing computable number should have a computable radius of convergence.

It turned out that Turing's notion of computability, in particular universal computability,
is quite robust in the sense that
it is equivalent to the recursive functions \cite{rogers1,odi:89}, abacus machines, or the usual
modern digital computer (given ``enough'' memory) based on the von Neumann architecture,
on which for instance this manuscript has been written and processed.

It is hardly questionable that Turing's model can be embedded
in physical space-time; at least in principle.
A discretization of physical space,
accompanied by deterministic evolution rules,
presents no conceptual challenge for a physical realization.
After all, Turing's conceptualization started from the intuitive
symbolic handling of the mathematical entities that every pupil is drilled to obey.
Even grown-up individuals arguably lack
an understanding of those rules imposed upon them and thus lack the semantics;
but this ignorance does not stop them from applying the syntax correctly, just as a Turing machine does.

There are two problems and two
features of any concrete technical realization of Turing machines.

(P1)
On all levels of physical realization, errors occur
due to malfunctioning of the apparatus.
This is unavoidable.
As a result, all our realistic models of computation must be
essentially probabilistic.

(P2)
From an operational perspective \cite{bridgman,bridgman52},
all physical resources are strictly finite and cannot be unbounded;
even not potentially unbounded
\cite{gandy1,gandy2}.

(F1)
It comes as no surprise that any embedding of a universal Turing machine,
and even more so less powerful finitistic computational concepts,
into a physical system results in physical undecidability.
In case of computational universality, this is due to a reduction to
the recursive unsolvability of the halting problem.
Ever after G\"odel's and Tarsky's destruction of the finitistic program of
Hilbert, Kronecker and others to find a finite set of axioms from which to derive
all mathematical truth, there have been attempts to translate these results into some
relevant physical form (e.g., see
Ref.~\cite{popper-50i,popper-50ii,moore,svozil-93,casti:94-onlimits_book,casti:96-onlimits}).

(F2)
The recursive undecidability of the
rule inference problem \cite{go-67} states that for any mechanistic agent
there exists a total
recursive function
such that the agent cannot infer this function.
In more physical terms,
there is no systematic way of finding a deterministic law from the
input-output analysis of a (universal) mechanistic physical system.

The undecidabilities resulting from (F1)\&(F2) should not be confused with
another mode of undecidability.
Complementarity is
a quantum mechanical feature also occurring in finite automata theory
\cite{e-f-moore,conway,svozil-93,schaller-96,dvur-pul-svo,cal-sv-yu,svozil-ql}
and generalized urn models \cite{wright:pent,wright},
two models having a common logical; i.e., propositional structure \cite{svozil-2001-eua}.

\section{Cantor's paradises and classical physics}

It is reasonable to require
from a ``useful'' theory of computation that any
capacity and feature of physical systems (interpretable as ``computing
machines'') should be reflected therein and
{\it vice versa.}
If one assumes some correspondence between
(physical) theory and physical systems
\cite{svozil-set,sv-aut-rev},
how does the continuum and its associated pandemonium of effects
(such as the Banach-Tarski paradox \cite{wagon1,pitowsky-82,svo-neufeld}; see also Ref.~\cite{siegel95})
fit into this picture?

\subsection{Computational correspondence between formal and physical entities}

According to the standard physics textbooks,
physical theory requires ``much'' richer structures than are
provided by universal Turing computability.
Physical theories such as (pre-quantum) mechanics \cite{goldstein}
and electrodynamics \cite{jackson} in various ways assume the continuum,
for example configuration space-time,
phase space, field observables and the like.
Even quantum mechanics is a theory based upon continuous space and time
as well as on a continuous wave function,
a fact which stimulated Einstein to remark (at the end of Ref.~\cite{ein1})
that maybe we should develop
quantum theory radically further into a purely discrete formalism.

Note that, with probability one, any element of the continuum is neither
Turing computable, nor algorithmic compressible; and thus random
\cite{chaitin3,calude:02}.
Thus, assuming that initial values of physical systems
are arbitrary elements ``drawn'' from some
``continuum urn''
amounts to assuming that in almost all cases they cannot be represented by
any constructive, computable method.
Worse yet, one has to assume the physical system has a capacity
associated with the axiom of choice in order to even make sure
that such a draw is possible.
Because how could one draw; i.e., select, an initial value,
whose representation cannot be represented in any conceivable algorithmic way?

These issues have become important for the conceptual foundation of chaos theory.
In the ``deterministic chaos'' scenario the deterministic equation of motion
appears to ``reveal'' the randomness; i.e.,
the algorithmically incompressible information
of the initial value \cite{shaw,schuster1,pit:96}.

Another issue is the question of the
preservation of computability in classical analysis,
the physical relevance of  Specker's theorems \cite{specker-ges,wang,kreisel},
as well as to the more recent constructions  by Pour-El and Richards \cite{pr1}
(cf. objections raised by Bridges
\cite{bridges1} and Penrose \cite{penrose:90});
see also Ref.~\cite{cal-sv-yu}.

\subsection{Infinity machines}

For the sake of exposing the problems associated with
continuum physics explicitly, an
oracle will be introduced
whose capacity exceeds  and outperforms any universal Turing machine.
Already Hermann Weyl
raised the question whether it is kinematically feasible
for a machine to carry out an {\em infinite} sequence of operations in
{\em finite} time;
see also
 Gr\"unbaum
\cite[p. 630]{gruenbaum:74}, Thomson \cite{thom:54}, Benacerraf \cite{benna:62},
Rucker \cite{rucker},
Pitowsky \cite{pit:90}, Earman and Norton \cite{ear-nor:93} and Hogarth
\cite{hogarth1,hogarth2},
as well as
Beth
 \cite[p. 492]{beth-59}
 and
L\'opez-Escobar
 \cite{le-91}, and the author \cite[pp. 24-27]{svozil-93} for related discussions.
Weyl writes \cite[p. 42]{weyl:49},
 \begin{quote}
 {\em
Yet, if the segment of length 1 really consists of infinitely many
sub-segments of length 1/2, 1/4, 1/8,
$\ldots$, as of `chopped-off' wholes, then it is incompatible with the
character of the infinite as the `incompletable' that Achilles should
have been able to traverse them all. If one admits this possibility,
then there is no reason why a machine should not be capable of
completing an infinite sequence of distinct acts of decision within a
finite amount of time; say, by supplying the first result after 1/2
minute, the second after another 1/4 minute, the third 1/8 minute later
than the second, etc. In this way it would be possible, provided the
receptive power of the brain would function similarly, to achieve a
traversal of all natural numbers and thereby a sure yes-or-no decision
regarding any existential question about natural numbers!   }
\end{quote}

The oracle's
 design is based upon
 a universal computer with ``squeezed'' cycle times of
 computation according to a geometric progression.
 The only difference
between universal computation
 and this type of
 oracle computation is the speed of execution.
 In order to achieve the limit,
 two time scales are introduced: the {\em intrinsic time $t$ scale
 of the process of computation}, which approaches infinity in finite
{\em
 extrinsic or proper time $\tau$ of some outside observer}.
The
time scales $\tau$ and $t$
 are related as follows.
\begin{description}
\item[$\bullet$]
The
 {\em proper time} $\tau$ measures the physical system time by clocks
 in a way similar to the usual operationalizations; whereas
\item[$\bullet$]
 a discrete {\it cycle time} $t=0,1,2,3,\ldots$ characterizes a sort of
 ``intrinsic'' time scale for a process  running on an otherwise
universal machine.
\item[$\bullet$]
 For some  unspecified
 reason we assume that this machine would allow us to
 ``squeeze'' its intrinsic time  $t$  with respect to the proper time
 $\tau$ by a geometric progression. Hence, for $k<1$, let
 any time cycle of $t$, if measured in terms of $\tau$, be squeezed by
 a factor of $k$ with respect to the foregoing time cycle
i.e.,
 \begin{eqnarray}
 \tau_0&=&0,\quad \tau_1=k ,\quad \tau_{t+1}-\tau_t =k(\tau_t
-\tau_{t-1}),
\\
 \tau_{t}&=&\sum_{n=0}^tk^n-1={k(k^t-1)\over k-1}\quad .
 \end{eqnarray}
Thus, in the limit
 of infinite cycle time $t\rightarrow \infty$, the proper time
 $\tau_\infty = k/(1-k)$ remains finite.
\end{description}
Note that for the oracle model introduced here
merely
dense space-time would be required.

As a consequence, certain tasks which lie beyond the domain of recursive
function theory  become computable and even tractable. For example, the
halting problem
and any problem codable into a halting problem would become solvable.
It would also be possible to produce an otherwise
 uncomputable and random output---equivalent to the tossing of a fair
 coin---such as Chaitin's halting probability \cite{chaitin3,calude:02}
 in finite proper time.

There is no commonly accepted physical principle which would forbid
such an oracle {\it a priori}.
One
might argue that such an oracle would require a geometric
energy increase  resulting in an infinite consumption of energy. Yet, no
currently accepted physical principle
excludes us from assuming that every geometric
decrease
in cycle time could be associated with a geometricaly decreasing progression in energy
consumption, at least up to some limiting (e.g., Planck) scale.

\section{Quantum oracles}

In the light of the quanta, the Church-Turing thesis, and in particular quantum recursion theory,
might have to be extended.
We first present an algorithmic form of a modified diagonalization procedure
in quantum mechanics due to the existence of fixed points of quantum information
\cite{deutsch91,svozil-paradox,ord-kieu-03}.
Then we shortly discuss quantum computation and mention recent proposals
extending the capacity of quantum computation beyond the Church-Turing barrier.

\subsection{Diagonalization method in quantum recursion theory}

Quantum bits can be physically represented by a coherent
superposition
of the two classical bit states denoted by $t$ and $f$.
The quantum bit states
\begin{equation}
x_{\alpha ,\beta}  =\alpha t+\beta f
\end{equation}
form a continuum, with
$ \vert \alpha \vert^2+\vert \beta \vert^2=1$, $\alpha ,\beta \in {\Bbb
C}$.

For the sake of
contradiction, consider a universal computer $C$ and  an arbitrary algorithm
$B(X)$ whose input is a string of symbols $X$.  Assume that there exists
a ``halting algorithm'' ${\tt HALT}$ which is able to decide whether $B$
terminates on $X$ or not.
The domain of ${\tt HALT}$  is the set of legal programs.
The range of ${\tt HALT}$ are classical bits (classical case) and quantum bits (quantum
mechanical case).

Using ${\tt HALT}(B(X))$ we shall construct another deterministic
computing agent $A$, which has as input any effective program $B$ and
which proceeds as follows:  Upon reading the program $B$ as input, $A$
makes a copy of it.  This can be readily achieved, since the program $B$
is presented to $A$ in some encoded form
$\ulcorner B\urcorner $,
i.e., as a string of
symbols.  In the next step, the agent uses the code
$\ulcorner B\urcorner $
 as input
string for $B$ itself; i.e., $A$ forms  $B(\ulcorner B\urcorner )$,
henceforth denoted by
$B(B)$.  The agent now hands $B(B)$ over to its subroutine ${\tt HALT}$.
Then, $A$ proceeds as follows:  if ${\tt HALT}(B(B))$ decides that
$B(B)$ halts, then the agent $A$ does not halt; this can for instance be
realized by an infinite {\tt DO}-loop; if ${\tt HALT}(B(B))$ decides
that $B(B)$ does {\em not} halt, then $A$ halts.

The agent $A$ will now be confronted with the following paradoxical
task:  take the own code as input and proceed.

\subsubsection{Classical case}
 Assume that $A$ is
restricted to classical bits of information.
To be more specific,
assume that ${\tt HALT}$ outputs the code of a classical bit as follows
($\uparrow$ and $\downarrow$ stands for divergence and convergence,
respectively):
\begin{equation}
{\tt HALT} ( B(X) ) =\left\{
 \begin{array}{l}
0 \mbox{ if } B(X) \uparrow
\\
1 \mbox{ if } B(X) \downarrow \\
\end{array}
 \right.
\quad .
\label{el:halt}
\end{equation}

Then, whenever $A(A)$
halts, ${\tt HALT}(A(A))$ outputs $1$ and forces $A(A)$ not to halt.
Conversely,
whenever $A(A)$ does not halt, then ${\tt HALT}(A(A))$ outputs $0$
and steers
$A(A)$ into the halting mode.  In both cases one arrives at a complete
contradiction.  Classically, this contradiction can only be consistently
avoided by assuming the nonexistence of $A$ and, since the only
nontrivial feature of $A$ is the use of the peculiar halting algorithm
${\tt HALT}$, the impossibility of any such halting algorithm.

\subsubsection{Quantum mechanical case}
As has been argued above, in quantum information theory
a quantum bit may be in a coherent
superposition
of the two classical states $t$ and $f$.
Due to this possibility of a coherent superposition of classical bit
states, the usual {\it reductio ad absurdum} argument breaks down.
Instead, diagonalization procedures in
quantum information theory yield quantum bit solutions which are fixed points
of the associated unitary operators.

In what follows it will be demonstrated how the task of the agent $A$
can be performed consistently if
$A$ is allowed to process quantum information.
To be more specific, assume that the output of the hypothetical
``halting algorithm'' is a quantum bit
\begin{equation}
{\tt HALT} ( B(X) ) = x_{\alpha , \beta}
\quad .
\end{equation}
We may think of   ${\tt HALT} ( B(X) )$ as a universal computer $C'$
simulating $C$ and containing a dedicated {\em halting bit}, which it
the output of $C'$
at every (discrete) time cycle. Initially (at time zero),
this halting bit is prepared to be a 50:50 mixture of the
classical halting and non-halting states $t$ and $f$; i.e.,
$x_{1/\sqrt{2} , 1/\sqrt{2} }$. If later $C'$ finds that $C$ converges
(diverges) on $B(X)$, then the halting bit of $C'$ is set to the
classical value $t$ ($f$).

The emergence of fixed points can be demonstrated by a simple example.
Agent $A$'s diagonalization task can be formalized as
follows. Consider for the moment the action of diagonalization  on the
classical bit states. (Since the quantum bit states are merely a coherent superposition
thereof, the action of diagonalization on quantum bits is straightforward.)
Diagonalization effectively transforms the classical bit value $t$ into $f$ and
{\it vice versa.}
Recall that in equation
(\ref{el:halt}),  the state
$t$ has been identified
 with the halting state and the state $f$
with the non-halting
state. Since the halting state and the non-halting state exclude each
other,
$f,t$ can be identified with orthonormal basis vectors  in a
twodimensional vector space. Thus, the standard basis of
Cartesian coordinates can be chosen for a representation of $t$ and $f$;
i.e.,
\begin{equation}
t  \equiv
\left(
\begin{array}{c}
1 \\
0
 \end{array}
\right)
\mbox{ and }
f \equiv
\left(
\begin{array}{c}
0 \\
1
 \end{array}
\right) \quad .
\end{equation}

 The evolution representing diagonalization (effectively, agent
$A$'s task) can be expressed by the unitary operator $D$ by
\begin{equation}
D t  =  f \mbox{ and }
D f  =  t\quad .
\end{equation}
Thus, $D$ acts essentially as a ${\tt not}$-gate.
In the above state basis, $D$ can be represented as follows:
\begin{equation}
D=
\left(
\begin{array}{cc}
0 & 1\\
1 & 0
\end{array}
\right) \quad .
\end{equation}
$ D $ will be called {\em diagonalization} operator, despite the fact
that the only nonvanishing components are off-diagonal.

As has been pointed out earlier,
quantum information theory allows a coherent superposition
$ x_{\alpha ,\beta}  =\alpha t+\beta f $
of the
classical bit states $t$ and $f$.
$D$ acts on classical bits. It
has a
fixed point at the classical bit state
\begin{equation}
x^\ast :=x_{ {1\over \sqrt{2}},{1\over \sqrt{2}} }  ={t+f\over \sqrt{2}}
\equiv
{1\over \sqrt{2}} \left(
\begin{array}{c}
1 \\
1
 \end{array}
\right) \quad .
\end{equation}
$x^\ast$
does not give rise to inconsistencies \cite{svozil-paradox}.
If agent $A$ hands over the fixed point state
$x ^\ast $ to the diagonalization
operator $D$, the same state
$x^\ast $ is recovered.
Stated differently, as long as the output of the ``halting
algorithm'' to input $A(A)$ is $x^\ast$, diagonalization does not
change it. Hence, even if the (classically) ``paradoxical'' construction
of diagonalization is maintained, quantum theory does not give rise to a
paradox, because the quantum range of solutions is larger than the
classical one.
Therefore,
standard proofs of the recursive unsolvability of the halting problem
do not apply if agent $A$ is allowed a quantum bit. The consequences for
quantum recursion theory are discussed below.

It should be noted, however, that the fixed point quantum bit ``solution''
to the above halting problem is of not much practical help.
In particular, if one is interested in the ``classical'' answer whether
or not $A(A)$ halts,  then one ultimately has to perform an
irreversible measurement
on the fixed point state. This  causes a state reduction into the
classical states corresponding to $t$ and $f$.
Any single measurement will yield an indeterministic result.
There is a 50:50 chance that
the fixed point state will be either in $t$ or $f$, since
$P_t(
x ^\ast)=
P_f(
x^\ast )= {1\over 2}$.
Thereby, classical undecidability is recovered.

Another, less abstract, application for quantum information theory is
the handling of inconsistent information in databases.
Thereby,
two contradicting classical bits of information
$t$ and
$f $ are resolved by the quantum bit
$x^\ast =
{(t+f)/ \sqrt{2}}$.
Throughout the rest of the computation the coherence is maintained.
After the processing, the result is obtained by an irreversible
measurement. The processing of quantum bits, however, would require an
exponential
space overhead on classical computers in classical bit base \cite{feynman}.
Thus, in order to remain tractable,
the corresponding quantum bits should be implemented on
truly quantum universal computers.

As far as problem solving is concerned, classical bits are not much of an
advance. If a classical information is required, then quantum bits are not
better than probabilistic knowledge. With regards to the question of
whether or not a computer halts, for
instance, the ``solution''
is equivalent to the throwing of a fair coin.

Therefore, the advance of quantum recursion theory over classical
recursion theory is not so much classical problem solving but {\em the
consistent representation of statements} which would give rise to
classical paradoxes.

The above argument used the continuity of classical bit states as compared to the
two classical bit states for a construction of fixed points of the
diagonalization operator. One could proceed a step further and allow
{\em nonclassical diagonalization procedures}. Thereby, one could allow
the entire range of twodimensional unitary transformations
\cite{murnaghan}
\begin{equation}
U_2(\omega ,\alpha ,\beta ,\varphi )=e^{-i\,\beta}\,
\left(
\begin{array}{cc}
{e^{i\,\alpha }}\,\cos \omega
&
{-e^{-i\,\varphi }}\,\sin \omega
\\
{e^{i\,\varphi }}\,\sin \omega
&
{e^{-i\,\alpha }}\,\cos \omega
 \end{array}
\right)
 \quad ,
\label{e:quid3}
\end{equation}
where $-\pi \le \beta ,\omega \le \pi$,
$-\, {\pi \over 2} \le  \alpha ,\varphi \le {\pi \over 2}$, to act on
the quantum bit.
A typical example of a nonclassical operation on a quantum bit is
the ``square root of not'' gate
($
\sqrt{{\tt not}}
\sqrt{{\tt not}} =D$)
\begin{equation}
\sqrt{{\tt not}} =
{1 \over 2}
\left(
\begin{array}{cc}
1+i&1-i
\\
1-i&1+i
 \end{array}
\right)
\quad .
\end{equation}
Not all these unitary transformations have eigenvectors
associated with eigenvalues $1$ and thus fixed points.
Indeed, it is not difficult to see that only
unitary transformations of the form
\begin{equation}
\begin{array}{l}
[U_2(\omega ,\alpha ,\beta ,\varphi )]^{-1}\,\mbox{diag}(1, e^{i\lambda
}) U_2(\omega ,\alpha ,\beta ,\varphi )=\\
\quad   \quad \quad
\left(
\begin{array}{cc}
{{\cos \omega }^2} + {e^{i\,\lambda }}\,{{\sin \omega }^2}&
{{{
{-1 + {e^{i\,\lambda
}}\over 2}
e^{-i\,\left(\alpha +\varphi \right) }}\,
\, \sin (2\,\omega )}} \\
{ -1 + {e^{i\,\lambda }}\over 2}
 {{{e^{i\,\left(\alpha
+\varphi \right) }}\,
 \sin
(2\,\omega )}}&
{e^{i\,\lambda }}\,{{\cos \omega }^2} + {{\sin
\omega }^2}
 \end{array}
\right)
 \end{array}
\end{equation}
have fixed points.

Applying nonclassical operations on quantum bits with no fixed points
\begin{equation}
\begin{array}{l}
[U_2(\omega ,\alpha ,\beta ,\varphi )]^{-1}\,\mbox{diag}( e^{i\mu } ,
e^{i\lambda }) U_2(\omega ,\alpha ,\beta ,\varphi )=\\
\quad   \quad \quad
\left(
\begin{array}{cc}
  {e^{i\,\mu }}\,{{\cos (\omega )}^2} +
     {e^{i\,\lambda }}\,{{\sin (\omega )}^2}&
    {{{e^{-i\,\left( \alpha  + p \right) }\over 2}}\,
         \left( {e^{i\,\lambda }} - {e^{i\,\mu }} \right) \,\sin
(2\,\omega )}
       \\
{{{e^{i\,\left( \alpha  + p \right) }\over 2}}\,
        \left( {e^{i\,\lambda }} - {e^{i\,\mu }}  \right) \,\sin
(2\,\omega )}
       &{e^{i\,\lambda }}\,{{\cos (\omega )}^2} +
     {e^{i\,\mu }}\,{{\sin (\omega )}^2}
 \end{array}
\right)
 \end{array}
\end{equation}
with $\mu ,\lambda \neq n\pi$, $n\in {\Bbb N}_0$ gives rise to
eigenvectors which are not fixed points, but which acquire nonvanishing
phases $\mu , \lambda$ in the generalized diagonalization process.

\subsection{Quantum computation}

First attempts to quantize Turing machines
\cite{deutsch}
failed to identify any possibilities to go beyond Turing computability.
Recently, two independent proposals by
Calude and Pavlov et al. \cite{2002-cal-pav,ad-ca-pa},
as well as by Kieu et al. \cite{kieu-02,kieu-02a}.
Both proposals are not just mere quantized extensions of Turing machines,
but attempt to utilize very specific features and capacities
of quantum systems.

The question as to what might be considered the ``essence'' of quantum computation,
and its possible advantages over classical computation, has been the topic of
numerous considerations, both from a physical
(e.g., Ref.~\cite{ekerj96,pres-97,pres-ln,nielsen-book,galindo-02,mermin-04,eisert-wolf-04})
as well as from a computer science
(e.g., Ref.~\cite{Gruska,benn:97,Ozhigov:1997,bbcmw-01,cleve-99,fortnov-03}) perspective.
One advantage of quantum algorithms over classical computation
is the possibility to spread out, process, analyse and extract information
in multipartite configurations in coherent superpositions of classical states.
This can be discussed
in terms of quantum state identification problems
based on a proper partitioning of mutually orthogonal sets of states
\cite{svozil-2005-ko}.

The question arises whether or not it is possible
to encode equibalanced decision problems
into quantum systems, so that a single invocation
of a filter used for state discrimination suffices to obtain the result.
Certain kinds of propositions about
quantum computers exist which do not correspond to any classical statement.
In quantum mechanics information
can be coded in entangled multipartite systems in such a way that
information about the single quanta is not useful for (and even makes impossible)
a decryption of the quantum computation.

Alas,  not all decision problems
have a proper encoding
into some quantum mechanical system
such that their resources (computation time, memory usage) is bound by some
criterion such as polynomiality or even finiteness.
One ``hard'' problem is the parity of a binary function
of $k>1$ binary arguments
\cite{Farhi-98,bbcmw-01,Miao-2001,orus-04,stadelhofer-05}:
It is only possible to go from $2^k$ classical queries down to $2^k/2$
quantum queries, thereby gaining a factor of 2.

Another example is a type of halting problem:
Alice presents Bob a black box with input and output interfaces.
Bob's task is to find out whether an arbitrary function of $k$ bits encoded in the black box
will ever output "0."
As this configuration could essentially get as worse as a {\em busy beaver} problem
\cite{rado,chaitin-bb},
the time it takes for Alice's box to ever output a "0" may grow faster than
any recursive  function of $k$.

Functional recursion
and iterations may
represent an additional burden on efficiency.
Recursions may require
a space overhead to keep track of the computational path,
in particular if the recursion depth cannot be coded efficiently.
From this point of view,
quantum implementations of the Ackermann or the Busy Beaver functions,
to give just two examples,
may even be less efficient than classical implementations,
where an effective waste management can get rid of many bits;
in particular in the presence of a computable radius of convergence.

\section{Dualistic transcendence}

It is an entirely different and open question whether or not the human
or animal mind can ``outperform'' any Turing machine.
Almost everybody, including eminent researchers, has an opinion on this matter,
but not very much empirical evidence has been accumulated.
For example, Kurt G{\"{o}}del believed in the capacity
of the human mind to comprehend mathematical truth beyond provability
\cite{kreisel-80,casti-jimmy}.

Why should the mind outpace Church-Turing computability?
The question is strongly related to the eternal issue of dualism and the
relation of body and soul (if any), of the mind and its brain,
and of Artificial Intelligence.
Instead of giving a detailed review of the related spiritual, religious
and philosophical \cite{descartes-meditation} discussions,
we refer to a recent theory based on neurophysiologic processes
by Sir John Eccles \cite{popper-eccles,eccles:papal}.

Even more speculitatively,
Jack Sarfatti allegedly (in vain) built an ``Eccles Telegraph'' in the form of an
electric typewriter directed by a stochastic physical process
which might be believed to allow communication with spiritual entities.
It may not be considered totally unreasonable to
base a theory of miracles \cite{frank,jung1}
on the spontaneous occurrence of stochastic processes
\cite{greenberger-svozil}
which individually may be interpreted to be ``meaningful,''
although their occurrence is statistically insignificant.

Dualism has acquired a new model metaphor in {\em virtual realities}
\cite{svozil-nat-acad} and the associated artistic expressions which
have come with it  (see, e.g., Refs.~\cite{simula,totalrecall,permutationcity,Matrix}).
We might even go as far as stating that we are the ``dead on vacation''
\cite{godard-aa}, or incarcerated in a Cartesian prison
(cf. Descartes' Meditation I,9 of Ref.~\cite{descartes-meditation}).
Some time ago, I had a dream. I was in an old, possibly medieval, castle.
I walked through it. At times I had the feeling that there
was something ``out there,'' something so
inconceivable hidden that it was impossible to recognize.
Then suddenly I realized that there was something ``inside the walls:''
another, dual,
castle, quite as spacious as the one I was walking in, formed by the
inner side of what one would otherwise consider masonry.
There was a small opening, and I glanced through; the inside looked
like a three-dimensional maze inhabited by dwarfs. The opening closed
again.

Computers are exactly such openings; doors of perception to hidden
universes.
In a computer-generated virtual environment the ``physical''
laws are deterministic and computable in the Church-Turing sense;
and yet this universe may not entirely be determined by the initial values and
the deterministic laws alone.
Dualism manifests itself in the two ``reality layers'' of the virtual reality
and the Beyond, as well as in the interface between them.
Through the interface, there can occur a steady flow of information back and forth
from and to the Beyond which is transcendental with respect to the
operational means available within the virtual reality.
Proofs of the recursive unsolvability of the halting problem or the rule inference problem,
for example,
break down due to the nonapplicability of
self-referential diagonal arguments in the transcendental Beyond.
This makes necessary a distinction between an extrinsic and an intrinsic representation
of the system \cite{svozil-94}.

\section{Verifiability}

Let us, in this final section, take up the thought expressed by Martin Davis in the first section;
and let us assume for a moment that some extraterrestrial visitors present
us a  device or ``oracle'' which
is purportedly capable to ``compute'' a non Church-Turing computable function.
In what follows we shall argue that we can do very little to verify such hilarious claims.
Indeed, this verification problem can be reduced to the induction problem,
which remains unsolved.

\subsection{Oracles in a black box}

However polished and suspicious the device looks,
for verification purposes
one may put it into a black box, whose only interfaces are
symbolic input and output devices, such as a keyboard and a digital display or printer.
The only important aspect of the black box is its input-output behaviour.

One (unrealistic) realization is a black box with an infinity machine stuffed into it.
The input and output ports of the infinity machine are directly connected to the input and
output interfaces of the black box.

The question we would like to clarify is this: how could observers by finite means know
that the black box represents an oracle doing something useful for us; in particular
computing a non Church-Turing computable function?

\subsection{Induction problem unsolved}

The question of verifiability of oracle computation can be related
to the question of how to differentiate a particular algorithm or more general
input-output behaviour from others.
In a very broad sense, this is the induction problem plaguing
inductive science from its very start.

Induction is ``bottom-up.'' It attempts to reconstruct certain postulated features
from events or the input-output performance of black boxes.
The induction problem, in particular algorithmic ways and methods
to derive certain outcomes or events from other (causally ``previous'')
events or outcomes via some kind of ``narratives'' such as physical theories,
still remains unsolved.
Indeed, in view of powerful formal incompleteness theorems, such as the halting problem,
the busy beaver function,
or the recursive unsolvability of the
rule inference problem, the induction problem is provable recursively unsolvable for
physical systems which can be reduced to, or at least contain, universal Turing machines.
The physical universe as we know it, appears to be of that kind
(cf. Refs.~\cite{svozil-93,svozil-unev}).

Deduction is of not much help with the oracle identification problem either.
It is ``top-down'' and postulates certain entities such as physical theories.
Those theories may just have been provided by another oracle,
they may be guesswork or just random pieces of data crap in a computer memory.
Deduction then derives empirical consequences from those theories.
But how could one possibly derive a non computable result if the only verifiable
oracles are merely Church-Turing computable?

\subsection{The conjecture on unverifiability beyond NP-completeness}

It is not totally unreasonable to speculate that
NP-completeness serves as a kind of boundary,
a demarcation line between operationally verifiable oracles
and nonverifiable ones.
For it makes no sense to consider propositions which cannot even be tractably verified.

\section{Outlook}

Presently the question of a proper formalization of the informal notion of
``algorithm'' seems to remain wide open.
With regards to discrete finite paper-and-pencil operations,
Church-Turing computability seems to be appropriate.
But if one takes into account physics,
in particular continuum mechanics and quantum physics,
the issues become less certain.
And if one is willing to include the full capacities of the human mind
with all its intuition and thoughtfulness,
any formalization appears highly speculative and inappropriate; at least for the time being,  but maybe forever.


\end{document}